\documentclass[prb,aps,twocolumn,amsmath,amssymb,floatfix,superscriptaddress]{revtex4}

\usepackage[dvips]{graphics}
\usepackage{color}
\usepackage{bbold}
\usepackage{physics}
\usepackage{soul}
\definecolor{dred}{rgb}{0.5,0.2,0}
\usepackage[colorlinks=true, citecolor=blue, urlcolor=blue]{hyperref}
\usepackage{tabularx}
\usepackage{graphicx} 
\date{\today}

\begin{document}

%\title{Comparative Theoretical Study of Spin and Charge Thermoelectric Transport in an Antiferromagnetic Triangular Ladder: Mechanisms for Enhanced Spin Thermoelectric Efficiency}
\title{Spin-Selective Thermoelectric Transport in a Triangular Spin Ladder}

\author{Ranjini Bhattacharya}

\email{ranjinibhattacharya@gmail.com}

\affiliation{Department of Condensed Matter and Materials Physics, S. N. Bose National Centre for Basic Sciences,
JD-Block, Sector III, Salt Lake, Kolkata 700098, India}

\author{Souvik Roy}

\email{souvikroy138@gmail.com}

\affiliation{School of Physical Sciences, National Institute of Science Education and Research, Jatni 752050, India}
\affiliation{Homi Bhabha National Institute, Training School Complex, Anushaktinagar, Mumbai 400094, India}

\begin{abstract}
% We present a detailed theoretical study of spin-dependent thermoelectric transport in a triangular ladder model with antiferromagnetic spin orientation. The unique topology of the ladder, combined with intrinsic spin-filtering effects, is shown to facilitate efficient spin-selective thermoelectric generation. By introducing distinct onsite energy modulations, namely, a binary spin-dependent modulation for up and down spins, and an aperiodic Aubry type modulation, we demonstrate how quasiperiodicity and spin asymmetry strongly affect the transport spectra. Furthermore, modulation of the hopping amplitudes enables control over multiple transport pathways, resulting in tunable spin-channel separation and enhanced thermoelectric response. Comprehensive analysis of the charge and spin-dependent transport coefficients reveals a substantial increase in the dimensionless thermoelectric figure of merit $ZT$ under optimized conditions. Remarkably, the spin $ZT$ consistently exceeds the charge counterpart, highlighting the crucial role of geometry and magnetic ordering in amplifying spin thermoelectric performance. These results establish a foundational framework for understanding and engineering spin-caloritronic functionality in low-dimensional architectures.

We theoretically investigate spin-resolved thermoelectric transport in a triangular ladder geometry hosting antiferromagnetic spin alignment, where lattice topology and magnetic ordering jointly enable highly efficient spin-selective energy conversion. The inherent geometric frustration of the ladder, together with intrinsic spin-filtering mechanisms, is shown to promote a pronounced separation between spin channels. Implementing spin-dependent onsite modulations, such as binary asymmetric potentials, induces pronounced spin splitting in the transmission spectrum, enabling controlled spin-selective transport and highlighting the role of lattice engineering in tailoring spin-dependent thermoelectric response. Additional control is achieved through modulation of the hopping amplitudes, which activates multiple transport pathways and allows fine tuning of spin-dependent conduction. A detailed evaluation of charge and spin thermoelectric coefficients reveals a strong enhancement of the thermoelectric performance, with the dimensionless figure of merit $ZT$ reaching large values in optimized parameter regimes. Notably, the spin figure of merit systematically surpasses its charge counterpart, underscoring the decisive role of lattice geometry and antiferromagnetic order in amplifying spin thermoelectric efficiency. Our findings provide a versatile theoretical platform for designing low-dimensional spin-caloritronic devices with enhanced functionality.
\end{abstract}

\maketitle
%\section{Introduction}

\section{\label{sec:intro}Introduction}
The steadily increasing global energy demand, together with mounting environmental concerns and the finite nature of fossil fuel resources, has intensified the search for alternative technologies capable of delivering high energy efficiency and sustainability. In this pursuit, thermoelectric (TE) systems have gained renewed interest due to their ability to convert thermal gradients directly into electrical power using solid-state mechanisms, without the need for mechanical motion or auxiliary working media. This intrinsic simplicity renders TE devices highly reliable, compact, and well suited for precise thermal regulation, positioning them as attractive candidates for waste-heat recovery and thermal management applications~\cite{TE1,TE2,TE3,TE4,TE5,TE6}. While the fundamental concepts underlying thermoelectricity date back to the discovery of the Seebeck and Peltier effects~\cite{seebeck1,peltier1}, the performance of existing thermoelectric materials remains far from optimal. Consequently, achieving a substantial enhancement of the dimensionless thermoelectric figure of merit, $ZT$, continues to be a central challenge and a key objective in the ongoing development of efficient thermoelectric materials and device architectures.

% The rapidly growing global demand for energy, together with the urgent concerns of climate change and the gradual exhaustion of fossil fuel reserves, has intensified the need for sustainable and high-efficiency energy technologies. In this context, thermoelectric (TE) devices have attracted considerable attention as they enable the direct conversion of waste heat into electrical energy without relying on moving components or working fluids. Owing to their structural simplicity, operational reliability, and capability for precise thermal control, TE systems offer distinct advantages for energy harvesting and thermal management applications~\cite{TE1,TE2,TE3,TE4,TE5,TE6}. Although the underlying principles of thermoelectricity originate from the classical Seebeck~\cite{seebeck1} and Peltier effects~\cite{peltier1}, the practical efficiency of contemporary TE devices remains limited. Consequently, improving the thermoelectric figure of merit ($ZT$) continues to be a fundamental and enduring challenge in the advancement of thermoelectric materials and devices.

Nanostructuring has provided transformative opportunities in thermoelectric research by overcoming many of the intrinsic limitations associated with bulk materials. Low-dimensional quantum systems, including quantum dots, carbon nanotubes, molecular junctions, and quantum wires~\cite{c13,c14,c15,c16}, have shown substantial enhancements in the thermoelectric figure of merit ($ZT$), enabling applications ranging from thermal rectification~\cite{rb1} to microscale refrigeration. Advances in single-molecule electronics and heterojunction engineering have further expanded the design space for thermoelectric devices, offering structurally versatile, electronically tunable, and environmentally sustainable platforms. These developments underscore the potential of low-dimensional systems~\cite{c10} as building blocks for scalable energy-harvesting technologies. Nevertheless, achieving consistently high thermoelectric efficiency while maintaining precise control over spin- and charge-dependent transport ~\cite{Ref18,Ref19,Ref20,Ref21} remains a central challenge. In this context, we introduce a triangular ladder model with tailored spin orientations at selected sites, combined with modulated onsite energies and hopping amplitudes, to achieve controlled spin-resolved transport. This framework provides a promising route toward designing highly efficient spin-selective thermoelectric devices with tunable transport characteristics.

% Nanotechnology has opened new avenues in thermoelectric research by enabling strategies that circumvent the intrinsic limitations of bulk materials. Low-dimensional quantum systems such as quantum dots, carbon nanotubes, molecular junctions, and quantum wires~\cite{c13,c14,c15,c16} have demonstrated significant enhancement of the thermoelectric figure of merit ($ZT$), motivating applications including thermal rectification~\cite{rb1} and microscale refrigeration. Recent advances in single-molecule electronics and hetero-junction engineering have further expanded thermoelectric architectures by introducing structurally flexible, electronically tunable, and environmentally sustainable platforms. These developments highlight the promise of low-dimensional systems~\cite{c10} for scalable energy-harvesting technologies. However, achieving stable high-efficiency performance together with precise control of transport properties remains a key challenge. To address this, we propose a novel framework that combines a triagular ladder with distinct spin orientation at specific sites. With onsite and hopping modulation, enables controlled spin-dependent transport and offers a viable route toward highly efficient spin-selective thermoelectric devices.

Motivated by these challenges, we propose a theoretical framework for spin-resolved transport based on a triangular ladder geometry. In this construction, the upper arm exclusively accommodates spin-up electrons, whereas the lower arm carries only spin-down electrons, effectively mapping the spin degree of freedom onto a spatial index. Introducing asymmetry between the two arms results in a controlled separation of the spin-resolved transmission channels, enabling spin-selective transport~\cite{spin,sp1,sp2} within a unified tight-binding formalism.

% Motivated by these unresolved issues, we introduce a theoretical transport framework based on a triangular ladder–like geometry with explicitly spin-resolved channels. The system is constructed such that the upper arm accommodates only spin-up electrons, while the lower arm supports only spin-down electrons, allowing the spin degree of freedom to be treated as an effective spatial index. Breaking the symmetry between the two arms leads to a controlled splitting of spin-resolved transmission functions, thereby enabling spin-selective transport within a unified tight-binding description.

From a thermoelectric standpoint, achieving high spin-dependent performance requires minimizing the overlap between spin-up and spin-down transmission spectra while introducing sharp energy-dependent features. Within the Landauer–Büttiker formalism, these conditions directly enhance the spin-resolved Seebeck coefficient \cite{spin2,s7,s8,s9,s10,s11,s12,s13,s14,s15,s16} via the energy derivatives of the transmission functions. We show that such criteria can be systematically engineered by adjusting the microscopic parameters of the triangular ladder. Specifically, we consider two symmetry-breaking strategies. In the first approach, a binary lattice is introduced in which the onsite energies of the two spin channels differ, producing spin-dependent shifts in the transmission spectra. In the second approach, the onsite energies are kept identical, while asymmetry is introduced through unequal hopping amplitudes along the upper and lower arms. Both mechanisms generate pronounced spin splitting in the transmission channels, creating favorable conditions for an enhanced spin thermoelectric response. We find in both the cases the spin TE coefficients surpass the charge counterpart, which is indeed a very desirable outcome.  

% From a thermoelectric perspective, high spin-dependent performance requires both a reduced overlap between spin-up and spin-down transmission spectra and the presence of sharp energy-dependent features. Within the Landauer–Büttiker framework, such conditions directly enhance the spin-dependent Seebeck coefficient through the energy derivatives of the transmission functions. We demonstrate that these requirements can be systematically satisfied by tuning the microscopic parameters of the ladder geometry. In particular, we analyze two distinct symmetry-breaking mechanisms. First, we introduce a binary lattice configuration in which the onsite energies associated with the two spin channels differ, leading to spin-dependent shifts in the transmission spectra. Second, keeping the onsite energies identical, we impose asymmetry through unequal hopping amplitudes along the upper and lower arms. Both approaches generate pronounced spin splitting in the transmission characteristics and yield favorable conditions for enhanced spin thermoelectric response.

This parameter-tunable transport framework offers a minimal yet highly flexible platform for controlling spin-dependent thermoelectric properties~\cite{rb2,rb3}. By allowing independent adjustment of spin-resolved transmission channels within a coherent quantum transport scheme, the model provides a practical pathway for optimizing spin Seebeck coefficients ~\cite{Ref22,Ref23,Ref24,Ref25} and achieving substantially enhanced values of the spin thermoelectric figure of merit.

% This parameter-controlled transport formalism provides a minimal yet versatile platform for engineering spin-dependent thermoelectric properties. By enabling independent manipulation of spin-resolved transmission channels within a coherent quantum transport framework, the proposed model establishes a concrete route toward optimizing spin thermoelectric coefficients and achieving large values of the spin figure of merit.

The paper is structured as follows. In Sec.~I, we outline the motivation and objectives of the study. Sec.~II introduces the triangular ladder model and the nonequilibrium Green’s function (NEGF) based transport formalism~\cite{c19,c20}. Sec.~III presents a detailed discussion of the results, and Sec.~IV concludes with a summary of the main findings and their implications.

% The paper is organized as follows. Section~I introduces the motivation and scope of the work. Section~II presents the model and NEGF-based transport formalism~\cite{c19,c20}. Section~III discusses the results, and Section~IV summarizes the main conclusions.

%\section{Quantum Design of Thermoelectric Transport: System Architecture and Hamiltonian Framework}
\section{Quantum framework for thermoelectric transport: Model, Hamiltonian and Theoretical Framework}

%\subsection{Quantum Transport in Ferromagnetic SSH Chains: Tight-Binding Hamiltonian Formalism}
\subsection{Model and Hamiltonian Formulation of Quantum Transport in Triangular Spin Ladder}

The source and drain electrodes are maintained at slightly different temperatures, $T+\Delta T/2$ and $T-\Delta T/2$, which generates a finite thermal bias $\Delta T$ across the junction. To ensure analytical clarity and experimental relevance, we restrict our analysis to the linear-response regime by assuming that $\Delta T$ is sufficiently small. The central anti-ferromagnetic (AFM) chain is composed of atomic sites
hosting localized magnetic moments, whose general orientation may be described by the polar angle $\theta_i$ and the azimuthal angle $\psi_i$. In the present study, we consider a simplified yet physically meaningful configuration in which all local
moments are rigidly aligned along the $+Z$ direction. This alignment stabilizes a uniform ferromagnetic phase with complete spin polarization and a net magnetization
oriented along $+Z$. Such an idealized setting, free from spin fluctuations and external perturbations, provides a transparent platform for examining how an
intrinsically spin-polarized background controls spin-dependent transport and the associated thermoelectric response.

% The source and drain electrodes are held at two slightly different temperatures, $T+\Delta T/2$ and $T-\Delta T/2$, thereby generating a finite thermal bias $\Delta T$ across the junction. For analytical transparency and experimental relevance, we confine our discussion to the linear-response regime by assuming $\Delta T$ is sufficiently small. The central ferromagnetic (FM) chain consists of atomic sites each carrying a localized magnetic moment, whose general orientation can be parameterized by the polar angle $\theta_i$ and azimuthal angle $\psi_i$. In this work, we focus on a simplified yet physically pertinent configuration in which all local moments are rigidly aligned along the $+Z$ direction. This alignment stabilizes a uniform ferromagnetic phase characterized by complete spin polarization and a net magnetization pointing along $+Z$. Such an idealized setting, devoid of spin fluctuations or external perturbations, provides a clean platform to examine how an intrinsically spin-polarized background governs the spin-dependent transport and thermoelectric response of the system.

\begin{figure}[ht]
{\centering \resizebox*{7.8cm}{4.5cm}{\includegraphics{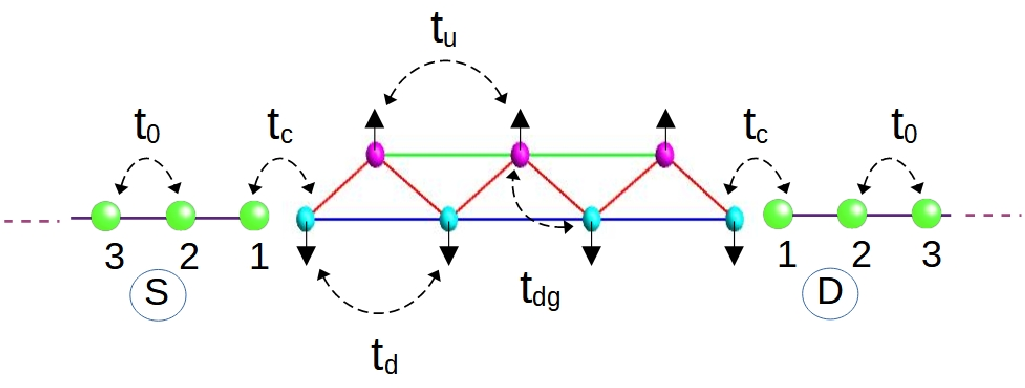}}\par}
%\caption{(\textcolor{red}{Color online).  
\caption{(Color online). The schematic illustration presents a triangular spin ladder connected to source and drain electrodes. This unique configuration opens a new frontier in thermoelectric transport by capturing interplay between diagonal hopping and onsite energy.}
\label{schematic}
\end{figure}

\begin{equation}
H= H_L + H_S + H_D + H_{tn}
\label{eq2}
\end{equation}
where, $H_L$, $H_S$, $H_D$, and $H_{tn}$ denote the Hamiltonians corresponding to the central chain, the source electrode, the drain electrode, and the tunneling
coupling between the electrodes and the SSH chain, respectively. Each term encapsulates a specific physical component of the setup, and their sum yields
a transparent decomposition of the total Hamiltonian. For completeness, the physical significance of each contribution is summarized below.

% where, $H_L$, $H_S$, $H_D$, and $H_{tn}$ represent the sub-Hamiltonians associated with the chain geometry, the source electrode, the drain electrode, and the tunneling connection between the electrodes and the central SSH chain, respectively. Each term highlights a distinct physical aspect of the system, and together they provide a systematic decomposition of the full Hamiltonian. For clarity and completeness, we briefly outline the role of each contribution in the following discussion.

% The term ${\bf h}_{N,i(M,i)} \cdot {\boldsymbol \sigma}$ governs the spin-dependent scattering processes within the system, where ${\boldsymbol \sigma}$ denotes the Pauli spin vector and $\sigma_z$ is assumed to be diagonal. Given our assumption that all local magnetic moments are uniformly aligned along the $+Z$ direction, the spin interaction term simplifies significantly. As a result, Eq.~\ref{eq5} reduces to a spin-diagonal ($4 \times 4$) matrix representation of the form $\mathrm{diag}\{h_i, -h_i, ~ h_i, -h_i\}$, effectively capturing the influence of ferromagnetic ordering on the electron spin dynamics.

The sub-Hamiltonians $H_S$ and $H_D$ appearing in Eq.~\eqref{eq2}, which describe the source and drain electrodes, respectively, are given by

% The sub-Hamiltonians $H_S$ and $H_D$ in Eq.~\eqref{eq2}, corresponding to the source and drain electrodes respectively, are expressed as follows:

\begin{equation}
H_S = \bf \sum_{n\leq -1} a^{\dagger}_{n}\epsilon_{0}a_{n}+\sum_{n\leq -1} \left(a^{\dagger}_{n}t_0a_{n-1}+a^{\dagger}_{n-1}t_0a_{n}\right)
\label{eq6}
\end{equation}
and
\begin{equation}
H_D = \bf \sum_{n\geq 1} b^{\dagger}_{n}\epsilon_{0}b_{n}+\sum_{n\geq 1} \left(b^{\dagger}_{n}t_0b_{n+1}+b^{\dagger}_{n+1}t_0b_{n}\right)
\label{eq7}
\end{equation}
Here, $\mathbf{t}_0$ and $\boldsymbol{\epsilon}_0$ denote the matrices $\mathrm{diag}\{t_0,t_0\}$ and $\mathrm{diag}\{\epsilon_0,\epsilon_0\}$,
respectively. The parameter $t_0$ represents the nearest-neighbor hopping amplitude, whereas $\epsilon_0$ is the on-site energy of the nonmagnetic
electrodes.

The tunneling contribution $H_{tn}$ is constructed in an analogous manner. As shown schematically in Fig.~\ref{schematic}, the source electrode is coupled
to the first site of the SSH chain, while the drain electrode is connected to the terminal site $m$. Under these considerations, the tunneling Hamiltonian
$H_{tn}$ takes the form

% Here, $\mathbf{t}_0$ and $\boldsymbol{\epsilon}_0$ represent the matrices diag$\{t_0, t_0\}$ and diag$\{\epsilon_0, \epsilon_0\}$, respectively. The parameter $t_0$ denotes the nearest-neighbor hopping integral, while $\epsilon_0$ corresponds to the on-site energy within the non-magnetic electrodes.

% The tunneling Hamiltonian $H_{tn}$ can be formulated analogously. As illustrated in Fig.~\ref{schematic}, the source electrode is connected to site $1$ of the SSH chain, while the drain is attached to the terminal site $m$. Accordingly, the sub-Hamiltonian $H_{tn}$ is expressed as
\begin{equation}
H_{tn} = \bf \left(a_{-1}^{\dagger}t_s c_1 + c_m^{\dagger}t_db_1 + h.c.\right)
\label{eqtn}
\end{equation}
where ${\bf t}_s=\mathrm{diag}\{t_s,t_s\}$ and ${\bf t}_d=\mathrm{diag}\{t_d,t_d\}$. The parameters $t_s$ and $t_d$ characterize the tunneling amplitudes between
the central conductor (chain) and the source (S) and drain (D) electrodes,
respectively.

% where ${\bf t_s} = \mathrm{diag}\{t_s, t_s\}$ and ${\bf t_d} = \mathrm{diag}\{t_d, t_d\}$. The parameters $t_s$ and $t_d$ quantify the coupling strengths between the conductor (chain) and the source (S) and drain (D) electrodes, respectively.

The triangular ladder system is described within a tight-binding framework,
where the spin degree of freedom is effectively mapped onto the spatial
structure of the ladder. The upper leg accommodates only spin-up electrons,
while the lower leg supports exclusively spin-down electrons. The total
Hamiltonian of the ladder can be written as
\begin{equation}
H_{\mathrm{L}} = H_{U,\uparrow} + H_{D,\downarrow} + H_{\triangle},
\end{equation}
where $H_{\uparrow}$ and $H_{\downarrow}$ correspond to the upper and lower legs,
respectively, and $H_{\triangle}$ accounts for the diagonal (triangular)
couplings between the two legs.

The Hamiltonian for the upper leg (spin-up channel) is given by
\begin{equation}
H_{U,\uparrow} =
\sum_{n} (\epsilon_{\uparrow}-h.\sigma)
c_{n,\uparrow}^{\dagger} c_{n,\uparrow}
- \sum_{n}
t_{u}
\left(
c_{n+1,\uparrow}^{\dagger} c_{n,\uparrow}
+ \mathrm{H.c.}
\right),
\end{equation}
where $c_{n,\uparrow}^{\dagger}$ ($c_{n,\uparrow}$) creates (annihilates) a
spin-up electron at site $n$ of the upper leg, $\epsilon_{n}^{\uparrow}$ denotes
the onsite energy, and $t_{n}^{\uparrow}$ is the nearest-neighbor hopping
amplitude along the upper arm.

Similarly, the lower leg (spin-down channel) is described by
\begin{equation}
H_{D,\downarrow} =
\sum_{n} (\epsilon_{\downarrow}+h.\sigma)
c_{n,\downarrow}^{\dagger} c_{n,\downarrow}
- \sum_{n}
t_{d}
\left(
c_{n+1,\downarrow}^{\dagger} c_{n,\downarrow}
+ \mathrm{H.c.}
\right),
\end{equation}
where $\epsilon_{n}^{\downarrow}$ and $t_{n}^{\downarrow}$ are the corresponding onsite energies and hopping integrals for the lower leg. The term $h \cdot { \sigma}$ describes the spin-dependent scattering processes in the system, where $ \sigma$ represents the Pauli spin vector and $\sigma_z$ is taken to be diagonal. Under the assumption that all localized magnetic moments are uniformly aligned along the $\pm Z$ direction, the spin interaction term simplifies considerably. 

The triangular geometry is introduced through diagonal inter-leg couplings,
which connect site $n$ of one leg to site $n+1$ of the opposite leg, and is
described by
\begin{equation}
H_{\triangle} =
- \sum_{n}
t_{dg}
\left(
c_{n+1,\uparrow}^{\dagger} c_{n,\downarrow}
+
c_{n+1,\downarrow}^{\dagger} c_{n,\uparrow}
+ \mathrm{H.c.}
\right),
\end{equation}
where $t_{dg}$ denotes the diagonal hopping amplitude that generates
the triangular ladder connectivity.

This construction ensures a strict spatial separation of spin channels, while
the diagonal couplings enable controlled spin-dependent interference pathways.
Such a geometry provides a natural platform for engineering spin-selective
quantum transport and enhanced thermoelectric response.

\subsection{Theoretical Framework}

% In this subsection, we outline the theoretical procedure for evaluating thermoelectric quantities, all of which fundamentally rely on spin-dependent transmission probabilities across the nano-junction. We begin by detailing the computation of the transmission function, followed by a systematic presentation of the subsequent steps required to extract the relevant thermoelectric parameters.
In this subsection, we outline the theoretical formalism used to compute the thermoelectric response of the system. Our analysis is based on the spin-resolved transmission probabilities of the nanojunction. We begin by defining the transmission function and subsequently describe, in a systematic manner, how the relevant thermoelectric coefficients are obtained from it.

% In this subsection, we present the theoretical framework employed to evaluate the thermoelectric response. The analysis is rooted in the spin-resolved transmission probabilities across the nanojunction. We first formulate the transmission function and then proceed with a step-by-step description of how the corresponding thermoelectric coefficients are extracted from it.

\vskip 0.2cm
\noindent
\emph{\textbf{Spin-Resolved Transmission Characteristics:}}
The spin-dependent transport properties of the nanojunction are analyzed within the Green’s function framework, which provides a rigorous and unified description of quantum transport while explicitly incorporating the influence of the electrodes. In this formalism, the retarded (advanced) Green’s function of the central region is given by
\begin{equation}
G^r=(G^a)^\dagger =[EI-H_R-\Sigma_S-\Sigma_D]^{-1},
\label{eq8}
\end{equation}
where $\Sigma_S$ and $\Sigma_D$ denote the self-energy matrices arising from the source and drain electrodes, respectively. These terms effectively encode the open-boundary conditions imposed by the contacts and account for the finite lifetime of electronic states due to their coupling with the leads. Here, $I$ is the identity matrix and $E$ represents the energy of the incident electron.

Once the Green’s functions are known, the spin-resolved transmission probabilities can be computed directly as
\begin{equation}
\tau^{\sigma\sigma^\prime}=\text{Tr}\!\left[\Gamma_S^\sigma G^r \Gamma_D^{\sigma^\prime} G^a\right],
\label{eq9}
\end{equation}
where $\tau^{\sigma\sigma^\prime}$ denotes the probability for an electron entering the system with spin $\sigma$ to exit with spin $\sigma^\prime$. The case $\sigma=\sigma^\prime$ corresponds to spin-conserving transport, whereas $\sigma\neq\sigma^\prime$ captures spin-flip scattering processes. The spin-dependent coupling matrices $\Gamma_S^\sigma$ and $\Gamma_D^{\sigma^\prime}$ quantify the hybridization strength between the central region and the respective electrodes and are obtained from the self-energy matrices via
\begin{equation}
\Gamma_{S(D)}^{\sigma\sigma^\prime}
=i\!\left[\Sigma_{S(D)}^{\sigma\sigma^\prime}
-\left(\Sigma_{S(D)}^{\sigma\sigma^\prime}\right)^\dagger\right].
\label{eq10}
\end{equation}

By incorporating both spin-preserving and spin-mixing channels, the effective transmission probabilities for the two spin sectors can be compactly expressed as
\begin{equation}
\tau^{\uparrow}=\tau^{\uparrow\uparrow}+ \tau^{\downarrow\uparrow},
\hspace{0.5cm}
\tau^{\downarrow}=\tau^{\downarrow\downarrow}+ \tau^{\uparrow\downarrow}.
\label{eq11}
\end{equation}

\vskip 0.2cm
\noindent
\emph{\textbf{Thermoelectric Response Functions:}}
To elucidate the thermoelectric performance of the system, we evaluate the spin-resolved Seebeck coefficient, electrical conductance, and electronic thermal conductance within the Landauer–Büttiker formalism. These key transport coefficients are expressed as
\begin{equation}
S^{\uparrow(\downarrow)}
=-\frac{1}{eT}\frac{L^{\uparrow(\downarrow)}_1}{L^{\uparrow(\downarrow)}_0},
\label{eq12}
\end{equation}
\begin{equation}
G^{\uparrow(\downarrow)}= e^2 L^{\uparrow(\downarrow)}_0,
\label{eq13}
\end{equation}
\begin{equation}
K^{\uparrow(\downarrow)}_{el}
=\frac{1}{T}\!\left(L^{\uparrow(\downarrow)}_2
-\frac{\left(L_{1}^{\uparrow(\downarrow)}\right)^2}{L^{\uparrow(\downarrow)}_0}\right).
\label{eq14}
\end{equation}

The Landauer integrals $L_n$ ($n=0,1,2$), which form the backbone of the thermoelectric analysis, are defined as~\cite{zt1,zt2}
\begin{equation}
L^{\uparrow(\downarrow)}_{n}
=-\frac{1}{h}\int \tau^{\uparrow(\downarrow)}
\left(\frac{\partial f(E)}{\partial E}\right)
\left(E-E_F\right)^{n}\, dE,
\label{eq15}
\end{equation}
where $f(E)$ is the Fermi–Dirac distribution function and $E_F$ denotes the equilibrium Fermi energy. Using these quantities, the charge and spin thermoelectric coefficients are constructed in a transparent manner. The charge and spin Seebeck coefficients are given by~\cite{zt1,zt2}
\begin{equation}
S_C=\frac{S^{\uparrow}+S^{\downarrow}}{2},
\hspace{0.5cm}
S_S=S^{\uparrow}-S^{\downarrow},
\label{eq16}
\end{equation}
while the corresponding electrical conductances read
\begin{equation}
G_C=G^{\uparrow}+G^{\downarrow},
\hspace{0.5cm}
G_S=G^{\uparrow}-G^{\downarrow}.
\label{eq17}
\end{equation}
The electronic thermal conductance takes the form
\begin{equation}
K_C=K_S=K^{\uparrow}+K^{\downarrow}=K.
\label{eq18}
\end{equation}

Owing to the nanoscale size of the system under consideration, the phonon-mediated thermal conductance $K_{ph}$ is expected to be negligibly small. We therefore neglect its contribution and, without loss of generality, identify the total thermal conductance with its electronic component, i.e., $K_e=K_C=K_S=K$.

Finally, the efficiency of thermoelectric energy conversion is quantified through the charge and spin figures of merit, defined as
\begin{equation}
Z_C T=\frac{S_C^2 G_C T}{K},
\hspace{0.5cm}
Z_S T=\frac{S_S^2 G_S T}{K}.
\label{eq19}
\end{equation}

\section{Numerical Results and Discussion}
\label{sec:results}
In this section, we present a comprehensive numerical study of the thermoelectric transport properties of the proposed model as shown in Fig.~\ref{schematic}, identifying parameter regimes that yield substantially enhanced thermoelectric performance, as reflected by large values of the figure of merit $ZT$. Our analysis demonstrates that the combined effects of hopping modulation and periodic or aperiodic onsite potentials play a central role in shaping the electronic transport characteristics and thermoelectric efficiency. By systematically tuning these parameters, we show how both charge and spin thermoelectric responses can be effectively optimized, providing concrete design guidelines for high-performance low-dimensional thermoelectric systems. Unless stated otherwise, the system size is fixed at $N=21$, the onsite energy of the leads is set to $\epsilon_0=0$, and the lead hopping strength is chosen as $t_0=2$. The coupling between the conductor and the source and drain electrodes is taken as $t_S=t_D=0.8$, ensuring symmetric and stable contact conditions. All calculations are performed at room temperature ($T=300$~K), and energies are expressed in units of electron-volts (eV). Parameter values specific to individual analyses are indicated where appropriate.

% In this section, we present a detailed numerical analysis of the thermoelectric transport characteristics of the proposed model, uncovering previously unexplored parameter regimes that yield remarkably enhanced thermoelectric performance, as evidenced by large values of $ZT$. Our results reveal that the delicate interplay among hopping amplitudes and periodic and aperiodic onsite potential modulations, plays a decisive role in governing both electronic transport and thermoelectric efficiency. By systematically tuning these parameters, we demonstrate how the system can be engineered to achieve optimized charge and spin thermoelectric responses, thereby offering new design principles for high-performance, low-dimensional thermoelectric materials. Unless otherwise specified, the total chain length is fixed at $N=21$, the on-site energy of leads $\epsilon_0$ is fixed at $0$, while the hopping integrals are chosen as $t_0 = 2$. The coupling strengths of the source and drain electrodes are set to $t_S = t_D = 0.8$, ensuring a balanced and stable coupling configuration. The temperature is maintained at $T = 300$~K, corresponding to standard room temperature conditions. Energy values are expressed in units of electron-volts (eV) to provide a uniform and standardized framework for analysis. Additional parameter values, when relevant, are provided in the corresponding sections to address specific aspects of the study.
\subsubsection*{Impact of upper-arm hopping $t_u$}
\subsection{Spectral Profile of the Transmission Function}

\begin{figure}[ht]
{\centering \resizebox*{7.5cm}{4cm}{\includegraphics{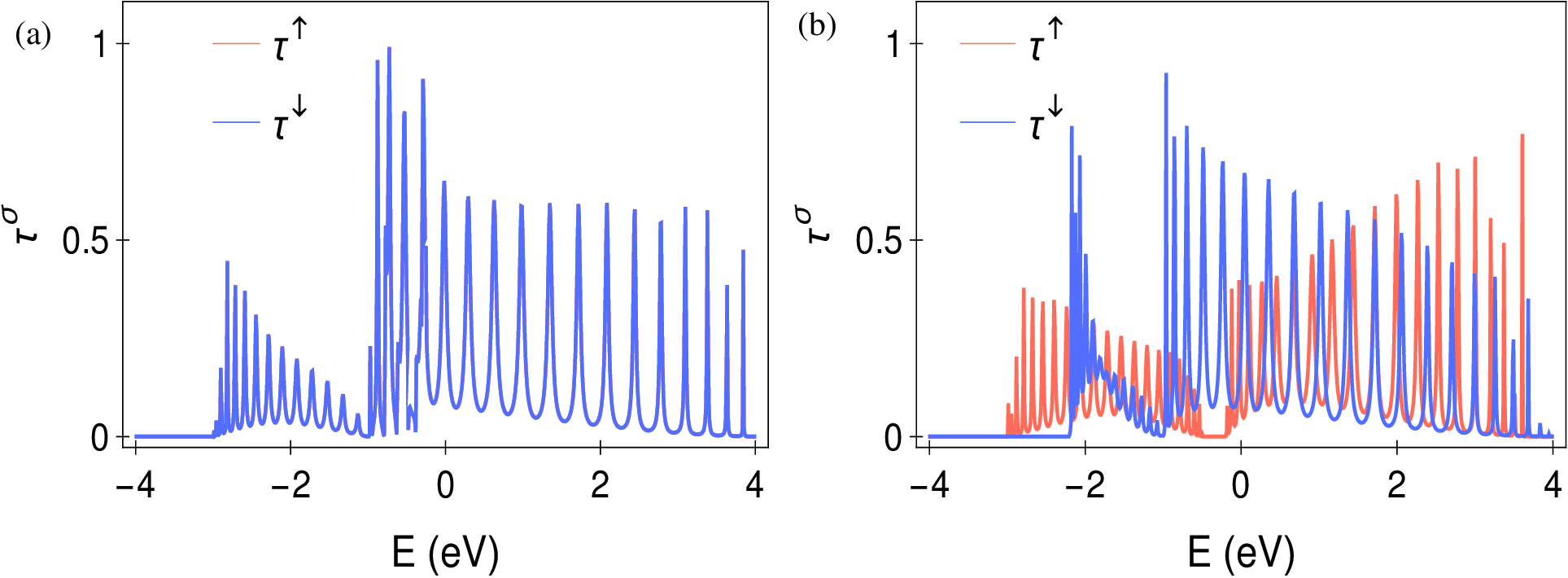}}\par}
\caption{(Color online). The plot illustrates the transmission function as a function of energy under various hopping configurations. Panel (a) displays fully spin overlapped transmission spectra, indicating the absence of spin-dependent separation when the hopping amplitude is $t_u = 1$. In contrast, panel (b) shows partial overlap between the spin-resolved channels, with $t_u=0.6$, clearly revealing the onset of spin-channel differentiation. This behavior reflects the subtle interplay between the distinct hopping pathways and their collective influence on spin-dependent transport characteristics.
}
\label{fig1}
\end{figure}

We begin our analysis by examining the energy-dependent transmission function 
$\mathcal{T}(E)$ under controlled variations of the hopping parameter 
$t_u$. When all hopping amplitudes are identical, i.e., $t_d = t_{dg} = t_u = 1$~eV  the transmission spectra for the up- and down-spin channels are completely overlapped, as shown in Fig.~\ref{fig1}(a), indicating the absence of spin-dependent transport. In contrast, when $t_d = t_{dg} = 1$~eV and $ t_u = 0.6$~eV a distinct partial separation between the up- and down-spin transmission spectra emerges [see Fig.~\ref{fig1}(b)]. This behavior clearly demonstrates that tuning $t_u$ serves as an effective means to control spin-selective transmission.

The origin of this spin-dependent asymmetry lies in the breakdown of geometric and hopping symmetry. When all hopping amplitudes are uniform, the structure remains symmetric for both spin orientations, resulting in identical transport characteristics. However, reducing $t_u$ disrupts this balance, leading to different effective environments for up- and down-spin electrons. The effect is more pronounced for the down-spin channel, which is more strongly influenced by $t_u$. Consequently, the resulting asymmetry in hopping strengths induces distinct spin-resolved transmission paths. Such control over spin-channel separation is particularly advantageous for optimizing spin thermoelectric efficiency, as minimal overlap between up- and down-spin transmissions is a prerequisite for achieving a large spin-dependent figure of merit. These results therefore highlight the strong potential of the present model for realizing efficient spin thermoelectric generators.

%\subsection{Seebeck Coefficient}
\subsection{Fermi-energy dependence of the Seebeck coefficient}

\begin{figure}[ht]
{\centering \resizebox*{7.5cm}{4cm}{\includegraphics{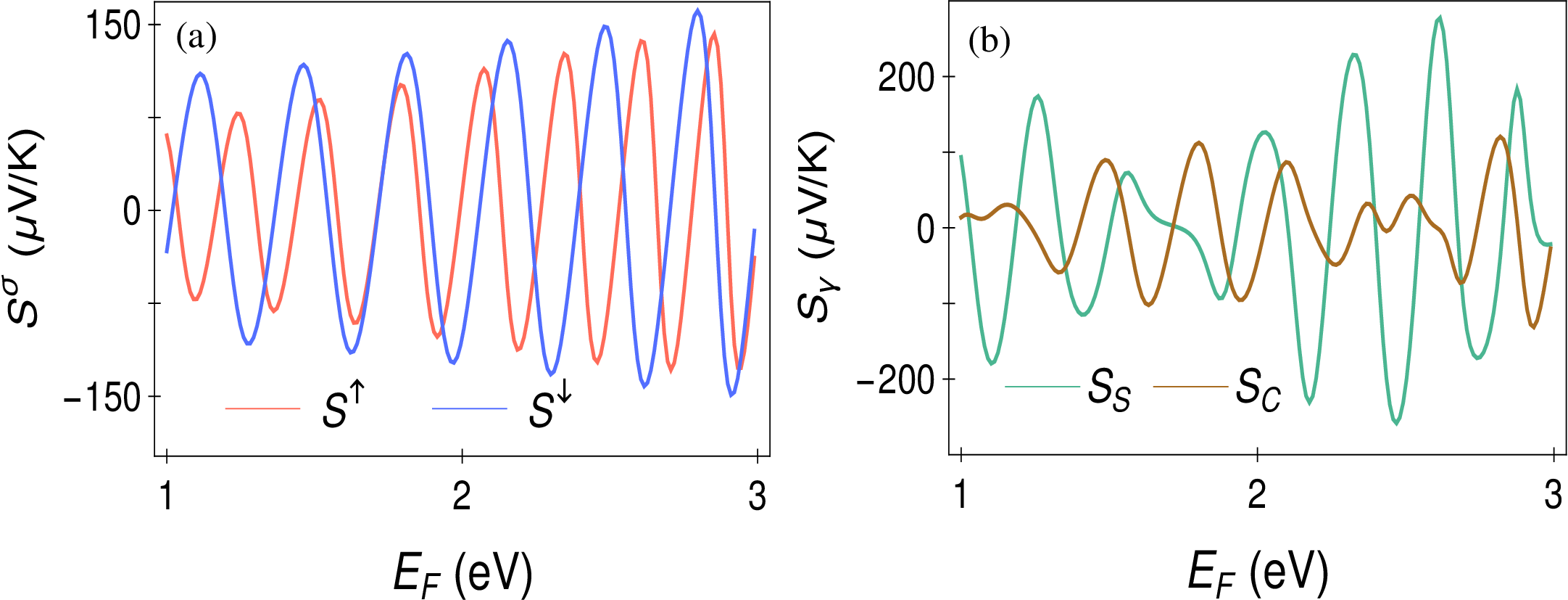}}\par}
\caption{(Color online). Seebeck coefficient as a function of Fermi energy is shown. Panel (a) displays the spin-resolved Seebeck response, showing distinct contributions from the up- and down-spin channels. Panel (b) presents the corresponding charge and spin Seebeck coefficients, thereby capturing the combined thermoelectric behavior arising from both transport channels.
}
\label{fig2}
\end{figure}

We next calculate the Seebeck coefficients for both spin channels. For this analysis, we adopt the same hopping parameters as in Fig.~\ref{fig1}(d), which are retained throughout the remainder of the study unless otherwise specified. The corresponding charge and spin Seebeck coefficients are then evaluated using Eq.~\ref{eq16}. To elucidate the origin and behavior of the Seebeck responses and their spin-resolved components, we analyze the central part of the Landauer integral, expressed as $\tau^\sigma(E)\left(\frac{\partial f}{\partial E}\right)(E-E_F)$, for each spin channel. The presence of the $(E-E_F)$
 term naturally allows $S^{\sigma}$ to assume both positive and negative values depending on the position of the Fermi energy $E_F$. As shown in Fig.~\ref{fig2}(a), the spin-resolved Seebeck coefficients exhibit a pronounced asymmetry in both magnitude and sign, with the up- and down-spin components contributing oppositely across a wide energy range. These distinct spin contributions combine to yield the total charge and spin Seebeck coefficients, $S_{\gamma}$ (with $\gamma = C,S$), displayed in Fig.~\ref{fig2}(b). Owing to the negative sign in Eq.~\ref{eq12}, the Seebeck coefficient varies inversely with the spin-resolved transmission function $\tau^\sigma$, hence, rising edges in $\tau^\sigma$,
 correspond to trailing features in $S^{\sigma}$, and vice versa. This interplay results in well-separated peaks in the Seebeck spectra, arising from the offset between the rising and falling edges of $\tau^{\uparrow}$ and $\tau^{\downarrow}$.
When $t_u$ differs from the other hopping parameters, the symmetry between the up- and down-spin transmission channels is broken, leading to distinct line shapes and energy separations in their transmission spectra. As shown in Fig.~\ref{fig1}(b), within the energy window $0$ to $4$ the crossing points of $\tau^{\uparrow}$ intersects $\tau^{\downarrow}$ coincide with opposite signs in $S^{\uparrow}$ and $S^{\downarrow}$ [see Fig.~\ref{fig2}(a)]. Owing to their nearly opposite characteristics over a wide energy range, a sizable spin Seebeck coefficient $S_S$ naturally emerges, surpassing the charge counterpart $S_C$, as evident in Fig.~\ref{fig2}(b). Quantitatively, the charge Seebeck coefficient reaches a maximum of approximately $120\,\mu$V/K, whereas the spin Seebeck coefficient attains nearly $270\,\mu$V/K,. A comprehensive analysis across the full Fermi energy range confirms that $S_S$ consistently dominates over $S_C$ Since the Seebeck coefficient enters quadratically in the numerator of the thermoelectric figure of merit $Z_{\gamma}T$ [Eq.~\ref{eq19}], its magnitude plays a decisive role in determining the overall thermoelectric efficiency. The observed large $S_S$ and comparatively smaller $S_C$ thus directly lead to a substantial enhancement of $Z_ST$ relative to $Z_CT$, as discussed in the subsequent subsection. These results clearly demonstrate that a controlled asymmetry between the spin-resolved transmission functions, together with their minimal overlap, can be strategically exploited to achieve a strong and tunable spin Seebeck response.

%\subsection{Electrical and thermal conductances}
\subsection{Fermi-energy dependence of electrical and thermal conductances}

\begin{figure}[ht]
{\centering \resizebox*{7.5cm}{4cm}{\includegraphics{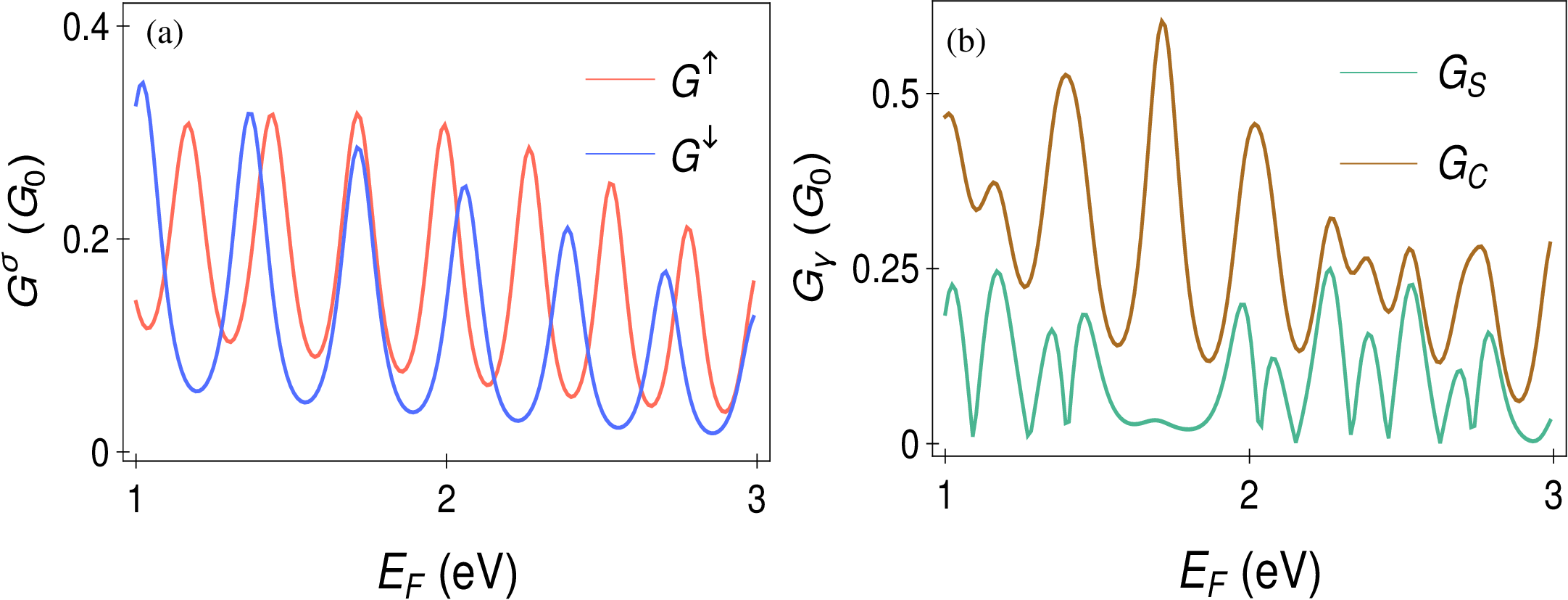}}\par}
\caption{(Color online). Panel (a) presents the spin-resolved electronic conductance as a function of Fermi energy, clearly illustrating the distinct contributions from the up- and down-spin carriers. Panel (b) compares the corresponding total charge and spin conductances, offering insight into the system’s thermoelectric response arising from spin-dependent transport channels.
}
\label{fig3}
\end{figure}

In this subsection, we investigate two additional thermoelectric quantities, electrical and thermal conductances which play a fundamental role in determining the efficiency of energy conversion in nanoscale systems. To ensure consistency and facilitate meaningful comparison, our analysis focuses on the same energy window as before, spanning 
$1$-$3$~eV where the overlap between the up- and down-spin transmission channels remains narrow. As expected from their physical definition, conductances are positive-definite, and our calculations confirm this behavior for both spin-resolved components. As shown in Fig.~\ref{fig3}(a), the spin-dependent electrical conductances $G^\uparrow$ and $G^\downarrow$ remain positive throughout the entire energy range, demonstrating robust and spin-selective transport. The asymmetry between these channels originates from choosing $t_u$ different from the other hopping parameters, which breaks the spin symmetry in the transmission spectra. The conductance profiles closely follow the features of the corresponding transmission functions, leading to multiple energy regions where 
$G^{\uparrow}$ and $G^{\downarrow}$ exhibit partial overlap. Since conductance is inherently positive, Eq.~\ref{eq17} provides a straightforward means of comparing the total charge and spin contributions. As depicted in Fig.~\ref{fig3}(b), the charge conductance $G_C$ consistently exceeds its spin counterpart $G_S$, with typical values of $G_C \approx 0.6~G_0$ and $G_S \approx 0.25~G_0$. This relatively higher $G_C$ contributes favorably to the charge figure of merit $Z_CT$. However, because the spin Seebeck coefficient $S_S$ is substantially larger than the charge Seebeck coefficient $S_C$, the overall thermoelectric efficiency in the spin channel remains superior. The enhanced $S_S$ thus plays a decisive role in elevating the spin-dependent figure of merit 
$Z_ST$ beyond its charge-based analogue, underscoring the potential of spin-polarized transport for high-efficiency thermoelectric applications.

\begin{figure}[ht]
{\centering \resizebox*{7.5cm}{5cm}{\includegraphics{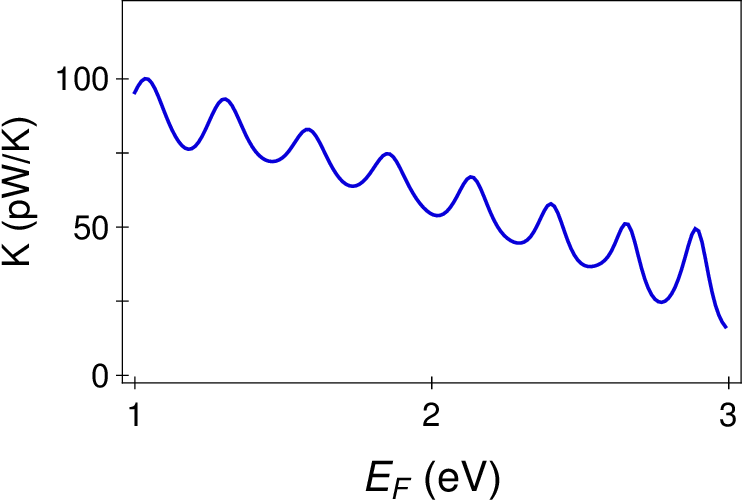}}\par}
\caption{(Color online). Thermal conductance as a function of Fermi energy, providing a comprehensive view of the system’s thermoelectric transport characteristics.
}
\label{fig4}
\end{figure}

The behavior of the thermal conductance is shown in Fig.~\ref{fig4}. Both charge and spin channels yield identical values of 
$K$, indicating that thermal conductance does not independently influence $Z_{C}T$ and $Z_{S}T$. Nevertheless, minimizing 
$K$ remains crucial for improving the overall thermoelectric efficiency, as it appears in the denominator of Eq.~\ref{eq19}. In our numerical analysis, the peak thermal conductance reaches approximately $100\,$pW/K. The variation of $K$ with Fermi energy closely follows the general trend of the charge conductance $G_C$, though the proportionality between the two is not exact across the full energy range. This departure from proportionality signals a clear violation of the Wiedemann–Franz (WF) law~\cite{wf1,wf2}, a feature frequently associated with enhanced thermoelectric performance in low-dimensional and nanoscale systems, and consistent with recent theoretical and experimental observations. It is important to emphasize that the present analysis considers only the electronic contribution to the thermal conductance. The phononic component has been deliberately omitted, as its effect is expected to be negligible due to the small system size under investigation.

\begin{figure}[ht]
{\centering \resizebox*{7.5cm}{5cm}{\includegraphics{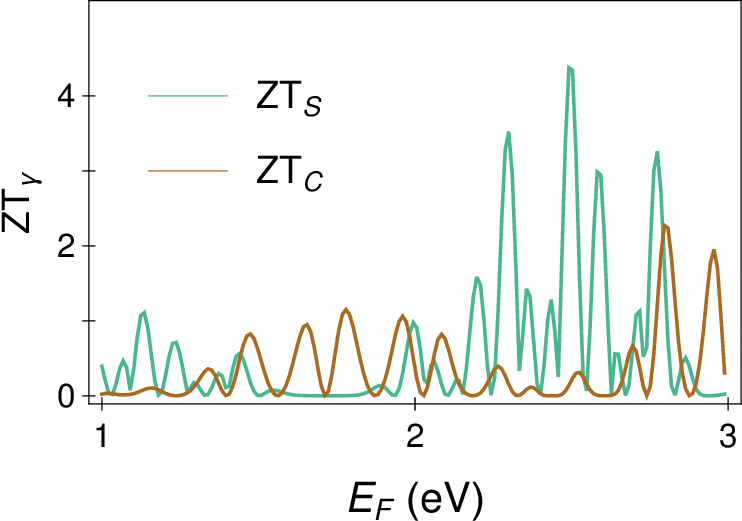}}\par}
\caption{(Color online). Thermoelectric figure of merit ($ZT$) as a function of Fermi energy. Both spin and charge $ZT$ are shown.
}
\label{fig5}
\end{figure}

%\subsection{Charge and spin figure of merits ($Z_CT$, $Z_ST$)}
\subsection{Tunable charge and spin figure of merits ($Z_CT$, $Z_ST$) with Fermi energy}

Building upon the detailed analysis of the thermoelectric quantities presented above, we now examine the variation of the charge and spin figures of merit, as shown in Fig.~\ref{fig5}. The green and brown curves correspond to the spin and charge responses, respectively. A pronounced and systematic enhancement of $Z_{S}T$ over $Z_{C}T$ is evident throughout the Fermi energy window ranging from $1$ to $3,$eV. Quantitatively, while the charge figure of merit $Z_{C}T$ reaches a maximum of approximately $2$, its spin counterpart, $Z_{S}T$, attains a substantially higher peak value of nearly $4$. This significant improvement in $Z_{S}T$ originates from the considerably larger spin Seebeck coefficient ($S_S$) compared to the charge Seebeck coefficient ($S_C$), a feature that directly stems from the intrinsic asymmetry between the spin-resolved transmission functions. As discussed earlier, the energy regions close to the crossing points of the up- and down-spin transmission spectra play a pivotal role in enhancing the spin thermoelectric response, as they naturally promote opposite spin contributions to the thermopower. The combined effect of transmission asymmetry and spin-channel separation thus provides a robust pathway toward optimizing spin-dependent thermoelectric efficiency. In the subsequent section, we further analyze how controlled variations in the hopping parameters govern this asymmetry, offering a deeper microscopic understanding of the tunability and design principles underlying efficient spin-selective thermoelectric devices based on such ladder-like geometries.

\subsubsection*{Impact of diagonal hopping}

In this subsection, we analyze the influence of the diagonal hopping amplitude under the condition $t_u \ne t_d$ on the thermoelectric performance of the system. Introducing a finite diagonal hopping term modifies the available conduction pathways and consequently the quantum interference patterns. Since the upper and lower arms of the ladder structure host opposite spin species, up spin in the upper arm and down spin in the lower, independent tuning of $t_u$, $t_d$ and $t_{dg}$ explicitly breaks the spin-channel symmetry. This asymmetry reshapes the transmission spectrum and can substantially enhance the spin-dependent thermoelectric coefficients. Physically, the simultaneous variation of arm and diagonal hoppings opens multiple interfering trajectories for spin transport. The resulting transmission asymmetry not only increases the magnitude of thermoelectric response but also shifts the Fermi-energy window where maximal values of $ZT$ are achieved. Thus, controlled manipulation of hopping parameters provides a powerful handle to optimize spin-selective thermoelectric efficiency in ladder-type quantum systems.

\begin{figure}[ht]
{\centering \resizebox*{7.5cm}{5cm}{\includegraphics{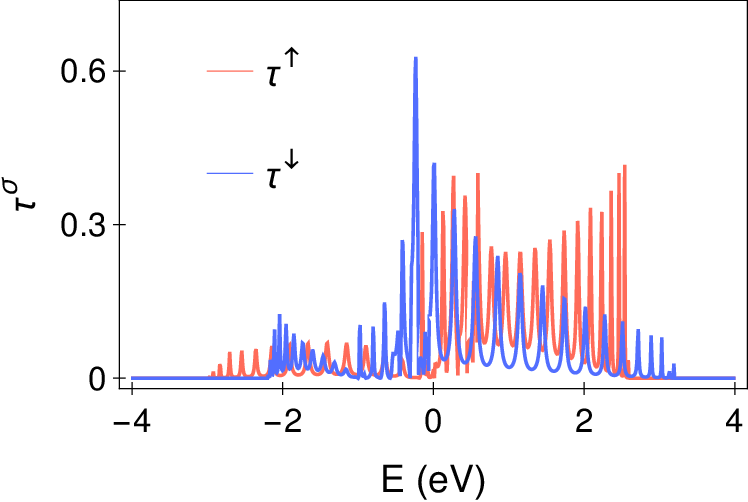}}\par}
\caption{(Color online). The plot illustrates the transmission function as a function of energy under variation of diagonal hopping. It shows partial overlap between the spin-resolved channels, with $t_u=0.6$, $t_d=1$ and $t_{dg}=0.5$. This behavior reflects the subtle interplay between the distinct hopping pathways and their collective influence on spin-dependent transport characteristics.
}
\label{fig11}
\end{figure}

We begin our analysis by examining the energy-dependent transmission function 
$\mathcal{T}(E)$ systematic variations of the diagonal hopping amplitude
$t_{dg}$. In Fig.~\ref{fig11}, we present the transmission spectra for $t_d = 1$, $ t_u = 0.6$ and $t_{dg} = 0.5$~eV. A clear departure from the behavior shown in
Fig.~\ref{fig1} is observed. In particular, the overlap between the spin-resolved
transmission channels increases relative to the uniform-hopping case, highlighting the
sensitivity of spin transport to the diagonal hopping strength. This trend indicates that
$t_{dg}$ serves as an effective tuning parameter for manipulating spin-dependent transport pathways.

The origin of the resulting asymmetry can be traced to the breakdown of geometric and
hopping symmetry. In the fully symmetric configuration (Fig.~\ref{fig1}), equal hopping
amplitudes ensure identical electronic environments for up- and down-spin carriers,
leading to spin-degenerate transmission. In contrast, reducing $t_u$ breaks
this balance by modifying the effective propagation channel for up-spin electrons.
Introducing or varying $t_{dg}$ further perturbs the interference conditions, producing a stronger distortion of the transmission line shape and enhancing the mixing between the spin-up and spin-down spectra.

Such controlled manipulation of spin-channel separation is particularly advantageous for
spin thermoelectric applications. Since a minimal overlap between the two spin channels is
a key requirement for achieving a large spin-dependent figure of merit, the present results
demonstrate that adjusting $t_{dg}$ provides a practical route toward
optimizing spin-selective thermoelectric performance. Overall, these findings underscore
the potential of the proposed ladder geometry as a platform for realizing efficient spin
thermoelectric generators.

%\subsection{Seebeck Coefficient}
\subsection{Evolution of the Seebeck coefficient across the Fermi energy}

\begin{figure}[ht]
{\centering \resizebox*{7.5cm}{5cm}{\includegraphics{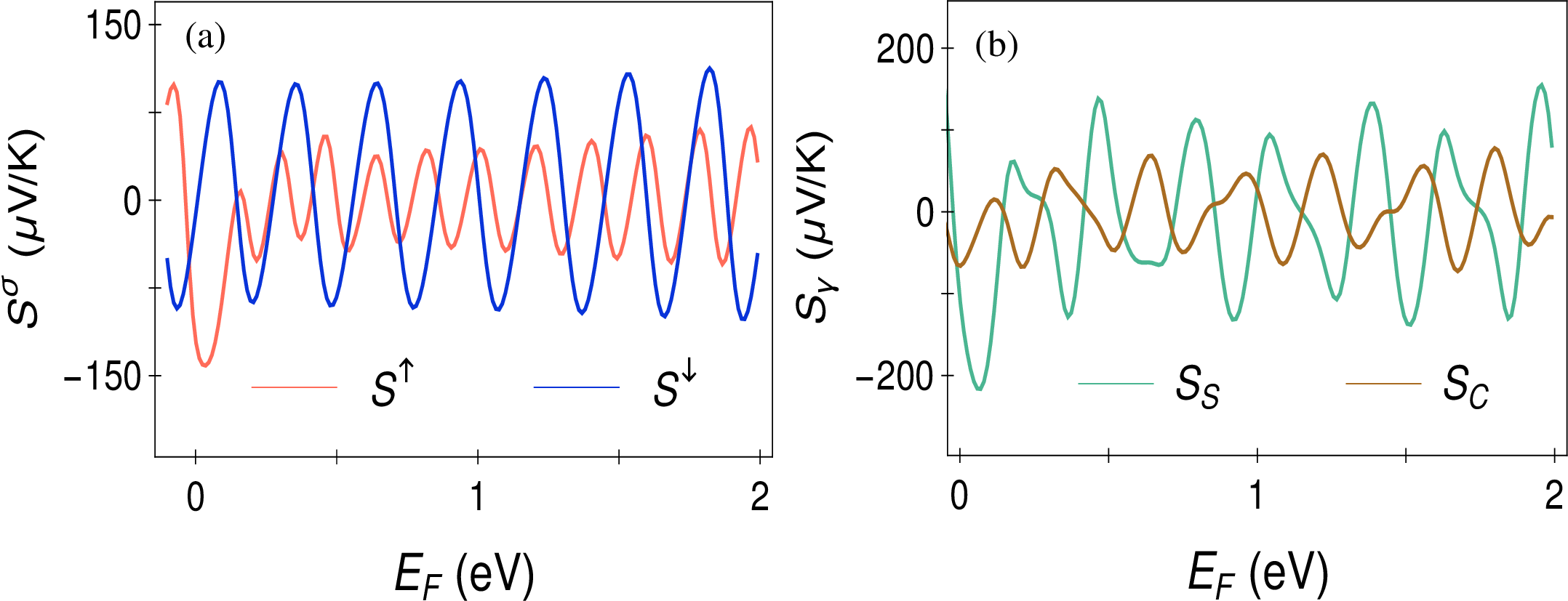}}\par}
\caption{(Color online).  Seebeck coefficient as a function of Fermi energy is shown. Panel (a) displays the spin-resolved Seebeck response, showing distinct contributions from the up- and down-spin channels. Panel (b) presents the corresponding charge and spin Seebeck coefficients, thereby capturing the combined thermoelectric behavior arising from both transport channels.
}
\label{fig12}
\end{figure}

We next calculate the Seebeck coefficients for both spin channels  Fig.~\ref{fig12}. For this analysis, we adopt the same hopping parameters as in Fig.~\ref{fig11}, which are retained throughout the remainder of the study unless otherwise specified. The corresponding charge and spin Seebeck coefficients are then evaluated using Eq.~\ref{eq16}. We calculate the Seebeck coefficient following similar procedure as mentioned in the previous subsections. Quantitatively, the charge Seebeck coefficient reaches a maximum of approximately $150\,\mu$V/K, whereas the spin Seebeck coefficient attains nearly $220\,\mu$V/K,. A comprehensive analysis across the full Fermi energy range confirms that $S_S$ consistently dominates over $S_C$ Since the Seebeck coefficient enters quadratically in the numerator of the thermoelectric figure of merit $Z_{\gamma}T$ [Eq.~\ref{eq19}], its magnitude plays a decisive role in determining the overall thermoelectric efficiency. The observed large $S_S$ and comparatively smaller $S_C$ thus directly lead to a substantial enhancement of $Z_ST$ relative to $Z_CT$, as discussed in the subsequent subsection. These results clearly demonstrate that a controlled asymmetry between the spin-resolved transmission functions, together with their minimal overlap, can be strategically exploited to achieve a strong and tunable spin Seebeck response.

We next evaluate the Seebeck coefficients for both spin channels using the same hopping parameters employed in Fig.~\ref{fig11}, which remain fixed throughout the remainder of this work unless stated otherwise. The charge and spin Seebeck coefficients are computed from Eq.~\ref{eq16}, following the procedure outlined in the preceding subsections. Quantitatively, the charge Seebeck coefficient $S_C$ exhibits a maximum value of approximately 150 $\mu$ V/K, whereas the spin Seebeck coefficient $S_S$ reaches nearly 220 $\mu$ V/K. A detailed examination across the entire Fermi-energy window reveals that $S_S$ consistently surpasses $S_C$, reflecting the underlying asymmetry between the spin-resolved transmission spectra. Since the Seebeck coefficient enters quadratically in the numerator of the thermoelectric figure of merit $Z\gamma T$ [Eq.~\ref{eq19}], its magnitude plays a crucial role in determining the overall thermoelectric performance. The substantially larger $S_S$, together with the comparatively smaller $S_C$, therefore directly accounts for the pronounced enhancement of $Z_S T$ relative to 
$Z_C T$, as discussed in the following subsection. These findings clearly demonstrate that controlled asymmetry in the spin-resolved transmission functions—combined with their reduced spectral overlap can be effectively leveraged to generate a strong and tunable spin Seebeck response. This establishes the present model as a promising platform for optimizing spin-dependent thermoelectric functionality.

%\subsection{Electrical and thermal conductances}
\subsection{Electrical and thermal conductances as functions of the Fermi energy}

\begin{figure}[ht]
{\centering \resizebox*{7.5cm}{5cm}{\includegraphics{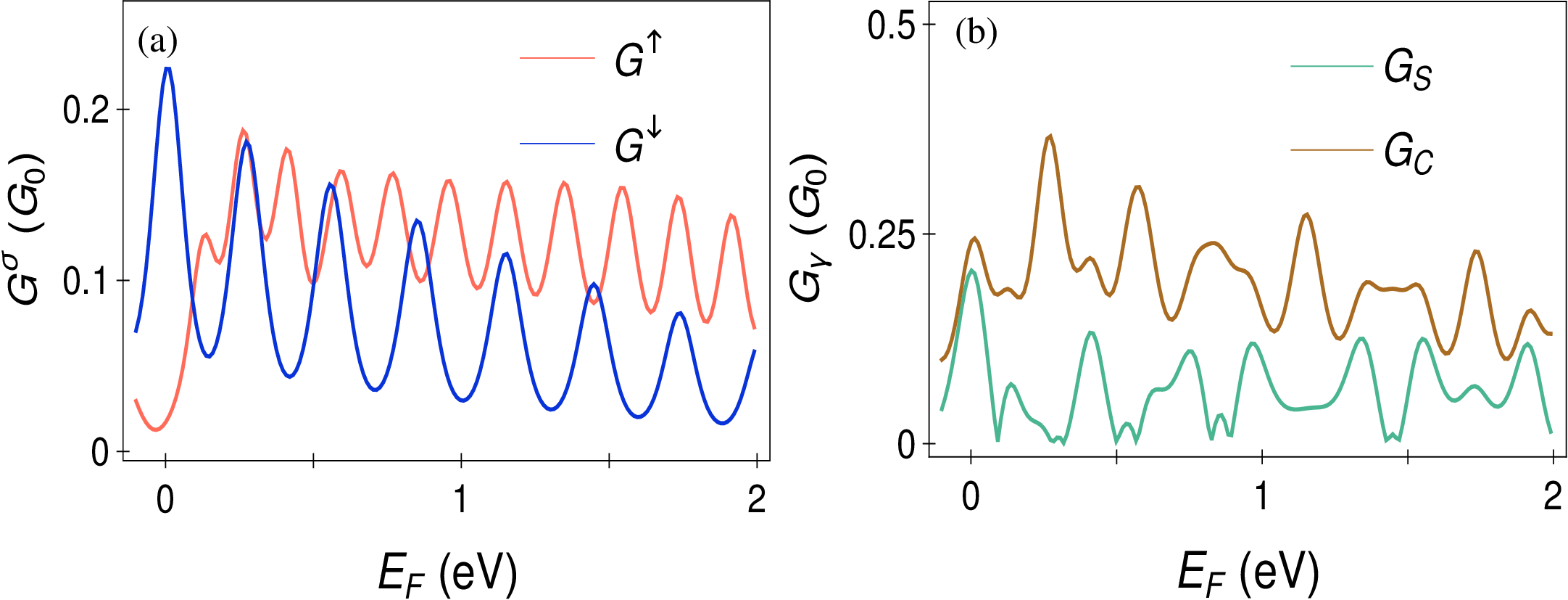}}\par}
\caption{(Color online). Panel (a) presents the spin-resolved electronic conductance as a function of Fermi energy, clearly illustrating the distinct contributions from the up- and down-spin carriers. Panel (b) compares the corresponding total charge and spin conductances, offering insight into the system’s thermoelectric response arising from spin-dependent transport channels.
}
\label{fig13}
\end{figure}

In this subsection, we analyze two key thermoelectric quantities—the electrical and thermal conductances,  Fig.~\ref{fig13} which are essential for evaluating the efficiency of energy conversion in nanoscale systems. As shown in Fig. 13(a), the spin-resolved electrical conductances, 
$G^{\uparrow}$ and $G^{\downarrow}$, remain positive over the entire energy range, signifying stable and spin-selective transport. The asymmetry between these two channels arises from the choice of distinct onsite energies, $\epsilon_{\uparrow} \ne \epsilon_{\uparrow}$, which explicitly breaks spin symmetry in the transmission spectra. The conductance profiles closely mirror the corresponding transmission characteristics, giving rise to multiple energy regions where $G^{\uparrow}$ and $G^{\downarrow}$ partially overlap. Since conductance is intrinsically positive, Eq. (14) provides a straightforward means of comparing the total charge and spin contributions. As shown in Fig. 13(b), the charge conductance $G_C$ consistently exceeds its spin counterpart $G_S$, with representative values of $G_C \approx 0.75~G_0$ and $G_C \approx 0.6~G_0$. The relatively larger $G_C$ contributes positively to the charge figure of merit $Z_CT$. However, because the spin Seebeck coefficient $S_S$ is considerably greater than the charge Seebeck coefficient $S_C$, the overall thermoelectric efficiency of the spin channel remains superior. The enhanced $S_S$ thus plays a decisive role in elevating the spin-dependent figure of merit $Z_ST$ beyond its charge-based analogue, underscoring the promise of spin-polarized transport for high-efficiency thermoelectric applications.

\begin{figure}[ht]
{\centering \resizebox*{7.5cm}{5cm}{\includegraphics{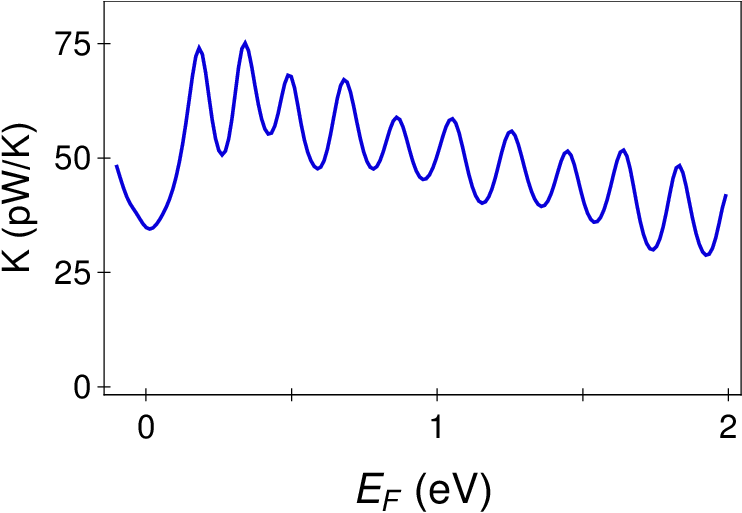}}\par}
\caption{(Color online). Thermal conductance as a function of Fermi energy, providing a comprehensive view of the system’s thermoelectric transport characteristics.
}
\label{fig14}
\end{figure}

The behavior of the thermal conductance is presented in  Fig.~\ref{fig14}. Both charge and spin channels yield identical thermal conductance values, indicating that $K$ does not directly differentiate $Z_CT$ and $Z_ST$. Nevertheless, minimizing $K$ remains crucial for optimizing overall thermoelectric performance, as it appears in the denominator of Eq. (16). In our calculations, the peak thermal conductance reaches approximately $150$~pW/K. It is important to note that the present analysis includes only the electronic contribution to $K$; the phononic part has been neglected, as its effect is expected to be minimal owing to the small system size under consideration.
\begin{figure}[ht]
{\centering \resizebox*{7.5cm}{5cm}{\includegraphics{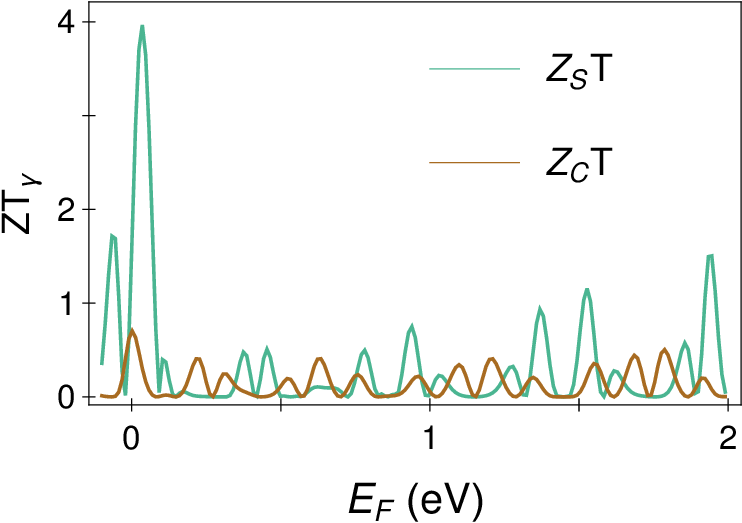}}\par}
\caption{(Color online). Thermoelectric figure of merit ($ZT$) as a function of Fermi energy. Both spin and charge $ZT$ are shown and the parameters used are consistent with those in Fig.~\ref{fig1} (d).
}
\label{fig15}
\end{figure}

%\begin{figure}[ht]
%{\centering \resizebox*{7.5cm}{5cm}{\includegraphics{tdiamax.eps}}\par}
%\caption{(Color online). Variation of maximum value of Thermoelectric figure of merit ($ZT$) as a function of diagonal hopping parameter is shown scanned over the whole Fermi energy window. Both spin and charge $ZT$ are shown.
%}
%\label{fig16}
%\end{figure}

\subsubsection*{Impact of binary type onsite energies}

In this subsection, we examine an alternative route to induce spin asymmetry. Instead of introducing different hopping amplitudes along the upper and lower arms, we consider a binary configuration of on-site energies for the up- and down-spin channels. This arrangement inherently breaks the spin symmetry and gives rise to distinct transport characteristics for the two spin species.

%\subsection{Spectral Profile of the Transmission Function}
\subsection{Energy-resolved transmission spectrum}

\begin{figure}[ht]
{\centering \resizebox*{7.5cm}{5cm}{\includegraphics{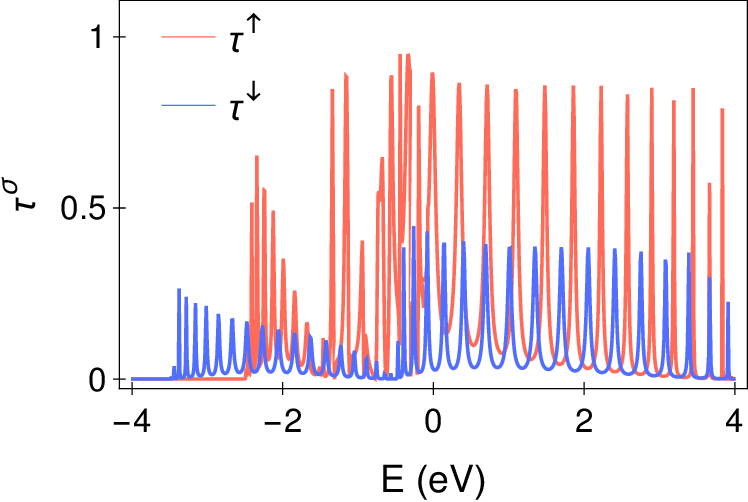}}\par}
\caption{(Color online). The plot illustrates the transmission function as a function of energy for different onsite energies. It reveals a partial overlap between the spin-resolved channels for $\epsilon_{\uparrow} = 0.5~\text{eV}$ and $\epsilon_{\downarrow} = -0.5~\text{eV}$, clearly indicating the emergence of spin-channel differentiation. This behavior highlights the delicate interplay between the distinct onsite energy landscapes and their combined influence on spin-dependent transport characteristics.
}
\label{fig6}
\end{figure}

We begin our analysis by examining the energy-dependent transmission function $\mathcal{T}(E)$,  Fig.~\ref{fig6} focusing on the effect of spin-dependent onsite energies. When all hopping amplitudes are identical, i.e., $t_d = t_{dg} = t_u = 1~\text{eV}$, and the onsite energies are set to zero, the up- and down-spin transmission spectra completely overlap (not shown), indicating the absence of spin-dependent transport. In contrast, introducing distinct onsite energies, $\epsilon_{\uparrow} = 0.5~\text{eV}$ and $\epsilon_{\downarrow} = -0.5~\text{eV}$, while keeping all hoppings equal, leads to a clear separation between the spin-resolved transmission spectra [see Fig.~6]. This behavior demonstrates that controlling the onsite energy asymmetry effectively modulates spin-selective transport.

The emergence of spin-dependent asymmetry originates from the breaking of the geometric and energetic equivalence between spin channels. In the uniform case, identical hoppings and onsite energies preserve full spin symmetry, resulting in degenerate transmission. However, the introduction of unequal onsite energies perturbs this balance, creating distinct local environments for up- and down-spin electrons and thus producing spin-resolved transmission pathways. Such control over spin-channel separation is crucial for enhancing spin thermoelectric performance, as reduced overlap between spin transmissions directly contributes to an increased spin-dependent figure of merit. These findings underscore the potential of this model in designing efficient spin-based thermoelectric generators.

%\subsection{Seebeck Coefficient}
\subsection{Seebeck response as a function of Fermi energy}

\begin{figure}[ht]
{\centering \resizebox*{7.5cm}{4cm}{\includegraphics{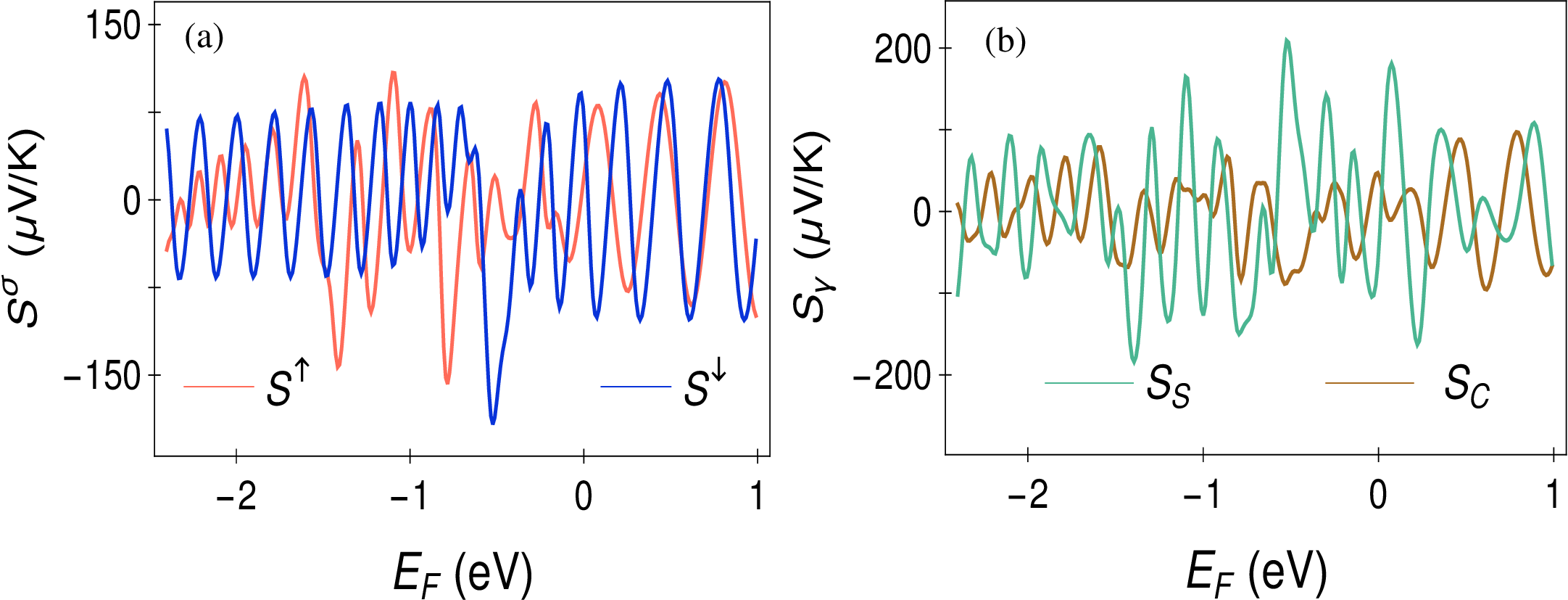}}\par}
\caption{(Color online). Seebeck coefficient as a function of Fermi energy is shown. Panel (a) displays the spin-resolved Seebeck response, showing distinct contributions from the up- and down-spin channels. Panel (b) presents the corresponding charge and spin Seebeck coefficients, thereby capturing the combined thermoelectric behavior arising from both transport channels.
}
\label{fig7}
\end{figure}

We next evaluate the Seebeck coefficients for both spin channels. In this analysis, we adopt the same set of hopping and onsite energy parameters as in  Fig.~\ref{fig7}, which are retained throughout this section unless otherwise stated. The corresponding charge and spin Seebeck coefficients are computed using Eq.~(13).

As discussed earlier, the presence of the $(E - E_F)$ term in Eq.~(13) naturally allows the spin-resolved Seebeck coefficient $S^{\sigma}$ to acquire both positive and negative values depending on the position of the Fermi energy $E_F$. As shown in  Fig.~\ref{fig7}(a), the spin-resolved Seebeck coefficients exhibit a distinct asymmetry in both magnitude and sign, with the up- and down-spin components contributing oppositely over a wide energy range. These contrasting spin contributions combine to yield the total charge and spin Seebeck coefficients, $S_{\gamma}$ (with $\gamma = C, S$), displayed in Fig.~7(b).

Due to the negative sign in Eq.~(9), the Seebeck coefficient varies inversely with the spin-resolved transmission function $\tau^{\sigma}$ rising edges in $\tau^{\sigma}$ correspond to trailing features in $S^{\sigma}$, and vice versa. This interplay leads to well-separated peaks in the Seebeck spectra, arising from the offset between the rising and falling edges of $\tau^{\uparrow}$ and $\tau^{\downarrow}$. For instance, as seen in Fig.~6, around $E \approx -1.5$~eV, $\tau^{\uparrow}$ increases while $\tau^{\downarrow}$ decreases, resulting in the opposite trend in $S^{\uparrow}$ and $S^{\downarrow}$ in  Fig.~\ref{fig7}(a).

Owing to their largely opposing characteristics across a broad energy window, a sizable spin Seebeck coefficient $S_S$ naturally emerges, surpassing its charge counterpart $S_C$, as evident from  Fig.~\ref{fig7}(b). Quantitatively, $S_C$ attains a maximum of about $100~\mu$V/K, whereas $S_S$ reaches nearly $200~\mu$V/K. A systematic inspection over the entire Fermi energy range confirms that $S_S$ consistently dominates $S_C$.

%\begin{figure}[ht]
%{\centering \resizebox*{7.5cm}{4cm}{\includegraphics{bi.eps}}\par}
%\caption{(Color online). Magnified version of  Fig.~\ref{fig7}, to understand how %transport asymmetry impacts Seebeck coefficient.
%}
%\label{fig8}
%\end{figure}

Since the Seebeck coefficient enters quadratically in the numerator of the thermoelectric figure of merit $Z_{\gamma}T$ [Eq.~(16)], its magnitude plays a decisive role in governing thermoelectric efficiency. The large $S_S$ and comparatively smaller $S_C$ thus directly contribute to the enhanced spin figure of merit $Z_S T$ relative to its charge counterpart $Z_C T$, as discussed in the next subsection. These findings clearly establish that a controlled asymmetry between the spin-resolved transmission channels combined with their minimal overlap can be effectively exploited to achieve a robust and tunable spin Seebeck response.

\subsection{Electrical and thermal conductances} 

\begin{figure}[ht]
{\centering \resizebox*{7.7cm}{4.5cm}{\includegraphics{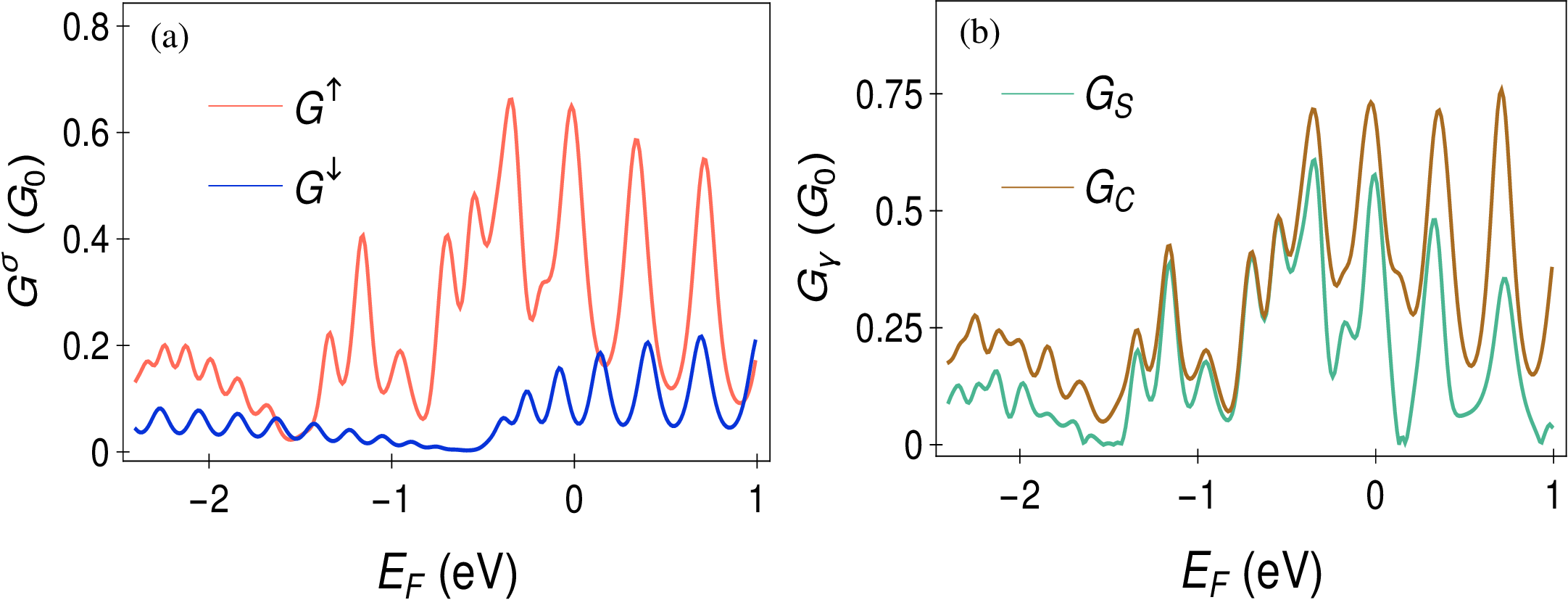}}\par}
\caption{(Color online). Panel (a) presents the spin-resolved electronic conductance as a function of Fermi energy, clearly illustrating the distinct contributions from the up- and down-spin carriers. Panel (b) compares the corresponding total charge and spin conductances, offering insight into the system’s thermoelectric response arising from spin-dependent transport channels.
}
\label{fig8}
\end{figure}

In this subsection, we analyze two key thermoelectric quantities—the electrical  Fig.~\ref{fig8} and thermal conductances which are essential for evaluating the efficiency of energy conversion in nanoscale systems. As shown in Fig. 8(a), the spin-resolved electrical conductances, 
$G^{\uparrow}$ and $G^{\downarrow}$, remain positive over the entire energy range, signifying stable and spin-selective transport. The asymmetry between these two channels arises from the choice of distinct onsite energies, $\epsilon_{\uparrow} \ne \epsilon_{\uparrow}$, which explicitly breaks spin symmetry in the transmission spectra. The conductance profiles closely mirror the corresponding transmission characteristics, giving rise to multiple energy regions where $G^{\uparrow}$ and $G^{\downarrow}$ partially overlap. Since conductance is intrinsically positive, Eq. (14) provides a straightforward means of comparing the total charge and spin contributions. As shown in Fig. 8(b), the charge conductance $G_C$ consistently exceeds its spin counterpart $G_S$, with representative values of $G_C \approx 0.75~G_0$ and $G_C \approx 0.6~G_0$. The relatively larger $G_C$ contributes positively to the charge figure of merit $Z_CT$. However, because the spin Seebeck coefficient $S_S$ is considerably greater than the charge Seebeck coefficient $S_C$, the overall thermoelectric efficiency of the spin channel remains superior. The enhanced $S_S$ thus plays a decisive role in elevating the spin-dependent figure of merit $Z_ST$ beyond its charge-based analogue, underscoring the promise of spin-polarized transport for high-efficiency thermoelectric applications. 

\begin{figure}[ht]
{\centering \resizebox*{7.5cm}{5cm}{\includegraphics{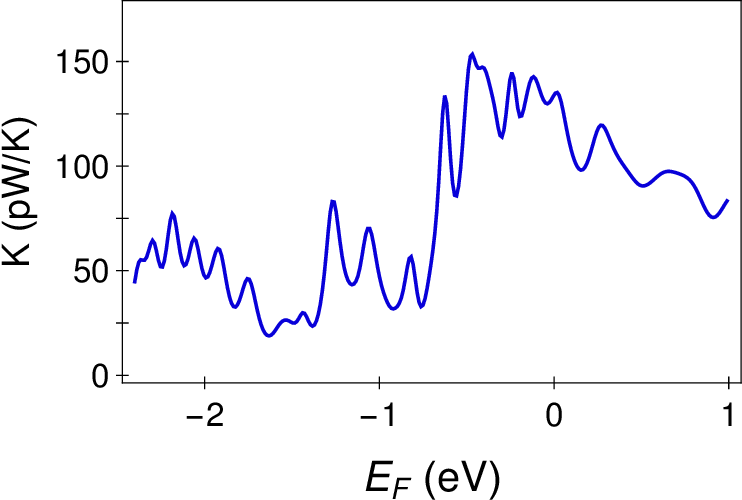}}\par}
\caption{(Color online). Thermal conductance as a function of Fermi energy, providing a comprehensive view of the system’s thermoelectric transport characteristics.
}
\label{fig9}
\end{figure}
The behavior of the thermal conductance is presented in  Fig.~\ref{fig9}. Both charge and spin channels yield identical thermal conductance values, indicating that $K$ does not directly differentiate $Z_CT$ and $Z_ST$. Nevertheless, minimizing $K$ remains crucial for optimizing overall thermoelectric performance, as it appears in the denominator of Eq. (16). In our calculations, the peak thermal conductance reaches approximately $150$~pW/K. It is important to note that the present analysis includes only the electronic contribution to $K$; the phononic part has been neglected, as its effect is expected to be minimal owing to the small system size under consideration.

%\subsection{Charge and spin figure of merits ($Z_CT$, $Z_ST$)} 
\subsection{Fermi Energy-resolved charge and spin figures of merit}
\begin{figure}[ht]
{\centering \resizebox*{7.5cm}{5cm}{\includegraphics{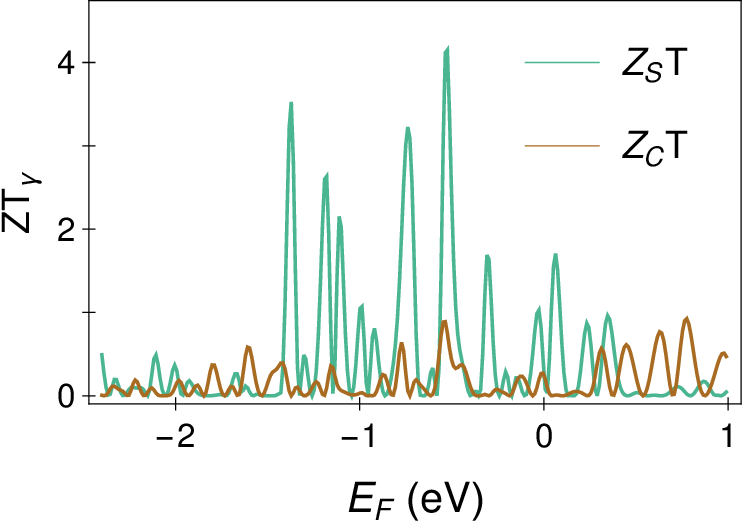}}\par}
%\caption{(\textcolor{red}{Color online).  
\caption{(Color online). This figure presents the phase diagram of thermoelectric parameters as functions of both $\alpha$ and $\beta$. Panel (a) and (d) display the variation of the maximum values of the spin and charge thermoelectric figure of merit ($ZT$), respectively. Panels (b) and (e) illustrate the corresponding maximum values of the spin and charge Seebeck coefficients. Finally, panels (c) and (f) show the maximum spin and charge conductance. Together, these panels provide a comprehensive overview of how the thermoelectric response for both spin and charge channels is modulated by the interplay between $\alpha$ and $\beta$.}
\label{fig10}
\end{figure}

Building upon the preceding analysis of thermoelectric transport, we now examine the variation of the charge and spin figures of merit, as shown in Fig. 16. The brown and green curves represent the charge ($Z_CT$) and spin ($Z_ST$) responses, respectively  Fig.~\ref{fig10}. A systematic enhancement of 
$Z_ST$ over $Z_CT$ is observed throughout the Fermi energy range from $-3$ eV to $1$ eV. Quantitatively, while $Z_CT$ reaches a maximum value of 
approximately 2, its spin counterpart attains nearly 4, demonstrating a substantial improvement in spin thermoelectric efficiency. This 
enhancement primarily arises from the markedly larger spin Seebeck coefficient ($S_S$) compared to its charge counterpart ($S_C$), which 
originates from the intrinsic asymmetry between the spin-resolved transmission functions. As discussed earlier, the energy regions near the 
crossing points of the up- and down-spin transmission spectra play a crucial role, as they induce opposite spin contributions to the thermopower 
and thereby amplify $S_S$. The combined influence of transmission asymmetry and spin-channel separation thus provides an effective mechanism for 
achieving enhanced and tunable spin-dependent thermoelectric performance.

\section{Conclusion}
\label{sec:conclusion}

In summary, we have developed a coherent theoretical framework to investigate spin-dependent thermoelectric transport in a triangular ladder geometry. By spatially segregating spin-up and spin-down channels and deliberately breaking the symmetry between the two arms, we achieve a controlled separation of spin-resolved transmission spectra. Within the NEGF formalism, we demonstrate that minimizing spectral overlap and enhancing energy asymmetry between opposite-spin transmission functions leads to a pronounced enhancement of the spin-dependent Seebeck coefficient. Two complementary symmetry-breaking routes, spin-dependent onsite energy modulation and asymmetric hopping along the ladder arms, are shown to provide robust and flexible control over spin-resolved transport without invoking external magnetic fields. Our analysis reveals a strong sensitivity of spin thermoelectric efficiency to microscopic transport parameters, enabling systematic optimization of the spin figure of merit. The proposed model simultaneously supports efficient spin filtering and enhanced spin caloritronic response, offering a minimal yet highly tunable platform for engineering spin-selective transport in low-dimensional systems. The results highlight the crucial role of geometry, symmetry breaking, and quantum coherence in achieving high-performance spin thermoelectric functionality and open avenues for future studies incorporating correlation effects and topological design principles.

\end{document}